\documentclass[sigplan,twocolumn,review]{acmart}

\usepackage{tikz}
\usepackage{amsmath}
\usepackage{url}
\usepackage{makecell}

\renewcommand\footnotetextcopyrightpermission[1]{}
\newcommand{\sysname}{MemServe}
\newcommand{\mempool}{MemPool}

\newcommand{\gs}{GS}
\newcommand{\GS}{GS}

\newcommand{\transInsert}{\texttt{transfer\_with\_insert}}
\newcommand{\transKV}{\texttt{transfer}}

\newcommand{\apiinsert}{\texttt{insert}}
\newcommand{\apimatch}{\texttt{match}}
\newcommand{\apidelete}{\texttt{delete}}
\newcommand{\apievict}{\texttt{evict}}

\newcommand{\ys}[1]{{{\color{red}#1}}{}}

\begin{document}

\date{}


\title{\sysname: Flexible Mem Pool for Building Disaggregated LLM Serving with Caching}

\author{
\rm
Cunchen Hu\textsuperscript{2,3},
Heyang Huang\textsuperscript{2,3},
Junhao Hu\textsuperscript{4},
Jiang Xu\textsuperscript{1},
Xusheng Chen\textsuperscript{1},
Tao Xie\textsuperscript{4}, \\
Chenxi Wang\textsuperscript{2,3},
Sa Wang\textsuperscript{2,3},
Yungang Bao\textsuperscript{2,3},
Ninghui Sun\textsuperscript{2,3},
Yizhou Shan\textsuperscript{1}
\break \\
\rm \large\textsuperscript{1}\textit{Huawei Cloud}, \textsuperscript{2}\textit{UCAS}
\textsuperscript{3}\textit{ICT, CAS}
\textsuperscript{4}\textit{Peking University}}

\pagestyle{plain}
\begin{abstract}

Large language model (LLM) serving has transformed from stateless to stateful systems, utilizing techniques like context caching and disaggregated inference.
These optimizations extend the lifespan and domain of the KV cache, necessitating a new architectural approach.
We present \sysname, a unified system that integrates both inter-request and intra-request optimizations.
\sysname\ introduces \mempool, an elastic memory pool managing distributed memory and KV caches across serving instances.
Using \mempool\ APIs, \sysname\ combines context caching with disaggregated inference for the first time, supported by a global scheduler that enhances cache reuse through a global prompt tree-based locality-aware policy.
Tests show that \sysname\ significantly improves job completion time and time-to-first-token.

\end{abstract}

\maketitle

\section{Introduction}

%

%
%
%
%

Large language models (LLMs) and their underlying transformer architecture have revolutionized AI, becoming foundational to many emerging applications and a crucial workload in data centers.
While high-quality models are essential, it is equally important to serve these models on a massive scale at a reasonably low cost. 
As a result, numerous approaches have been proposed to enhance the cost-efficiency of LLM serving, such as context caching~\cite{sglang,pensieve}, disaggregated inference~\cite{zhong2024distserve,patel2023splitwise}, and sequence parallelism~\cite{infiniteLLM}.

As a result, LLM serving has evolved from a stateless to a stateful system, leveraging dependencies inherent in inference requests. These dependency-exploiting techniques can be classified into two types: inter-request and intra-request.
Inter-request techniques exploit dependencies across requests. 
The notable one is context caching~\cite{sglang}, which reuses the KV cache for requests that share the same prompt prefix, thereby speeding up the prefill phase.
Intra-request techniques, on the other hand, exploit dependencies within a single request.
Two prominent examples are disaggregated inference, which splits a request into two sub-requests for better scheduling~\cite{patel2023splitwise}, and sequence parallelism, which divides a request into multiple sub-requests to distribute load~\cite{infiniteLLM}. 

A common theme in these dependency-exploiting techniques is that they require novel logic to manage and transfer the KV cache, which is the intermediate data produced during LLM inference.
Inter-request methods preserve the KV cache across requests, extending its lifetime from a single request to potentially infinite. Intra-request methods transfer the KV cache across multiple inference instances, extending its domain from a single instance to distributed instances.
However, deploying a stateful LLM serving system with these optimizations is challenging due to conflicting or missing mechanisms for managing the LLM's intermediate KV cache data. We have identified two key problems.

The first problem is that LLM serving systems cannot
simultaneously apply any existing inter-request and intra-request dependency-exploiting optimizations.
Current context caching (inter-request) methods are designed without considering intra-request scenarios.
As a result, disaggregated inference (intra-request) cannot benefit from context caching because it lacks the mechanisms to utilize the KV cache from decode back to prefill instances for future reuse. Similarly, sequence parallelism distributes the KV cache across multiple instances and lacks the mechanisms and algorithms needed to preserve and reuse it.
%
%
This issue arises because intra-request techniques break a tightly coupled request into multiple loosely coupled sub-requests, complicating KV cache management in a distributed setting.

The second problem is that LLM serving systems lack a holistic, top-down design to effectively utilize existing inter-request techniques.
Context caching benefits from reusing historical KV cache by running requests that share a common prefix in the same serving instance.
However, current LLM serving systems schedule requests across multiple serving instances based on load or session IDs, which fails to maximize KV cache reuse across sessions.

These issues arise because existing LLM serving systems are built on the assumption that the KV cache is merely intermediate data scoped to a single request on a single instance.
With emerging dependency-exploiting techniques, the lifespan of the KV cache has been extended, and its management has expanded to a distributed setup. 
This paradigm shift calls for a fundamental rethinking of LLM serving architectures.



In this work, we propose
\textbf{M}emory-\textbf{e}nhanced \textbf{m}odel Serving, or \sysname, to handle 
inter-request and intra-request optimizations within a unified system.
To tackle the challenges of managing the KV cache across distributed instances, \sysname\ introduces an elastic memory pool, or \mempool, which is a substrate for managing all cluster memory, including CPU DRAM and GPU HBM.
\mempool\ offers a rich set of APIs for managing distributed memory and KV cache.
Utilizing these APIs, \sysname\ implements context caching over standard prefill-decode-colocated (PD-colocated) instances~\cite{sglang} and disaggregated inference~\cite{zhong2024distserve,tetriserve-2024}.
Moreover, \sysname\ enhances disaggregated inference with context caching, reaping both benefits.
Finally, to maximize KV cache reuse, \sysname\ employs a global scheduler that incorporates a locality-aware policy using novel global prompt trees for best-effort routing.

The \mempool\ is a core component of \sysname, providing three types of APIs: memory, indexing, and distributed data transfer.
It runs within each inference instance, managing all local memory with a fixed-size memory allocator.
The indexing APIs are crucial for building context caching. \mempool\ uses an internal index to map prompt tokens to the KV cache, managing both the active KV cache for ongoing requests and the historical KV cache retained after requests are completed.
The \mempool\ offers a simple data transfer API that abstracts three heterogeneities: parallelism, network, and memory medium.
As a unified platform,
\mempool\ supports all known inter-request and intra-request optimizations as well as any combinations (see Figure~\ref{fig-mempool-use-cases}).

\sysname\ bridges the gap between context caching (inter-request) and disaggregated inference (intra-request) in four steps using \mempool\ APIs:
(a) we first use a distributed API to reproduce disaggregated inference,
(b) we then add caching to prefill-only instances using index APIs,
(c) we apply the same caching to decode-only instances,
(d) finally we enable decode-to-prefill data transfer, as illustrated in Figure~\ref{tbl-mempool-pd-caching-design}.
However, it is challenging to hit two birds with one stone.
We observed increasing overheads due to naive discrete memory layouts and point-to-point network primitives from existing AI network stacks.
To address this, \sysname\ proposes co-optimizing memory layout and network transfer using huge pages.

We implement \mempool\ and global scheduler from scratch, 5.6K SLOC in Python and
1.6K SLOC in C++
We modify vLLM~\cite{vllm-sosp23} to build context caching with disaggregated inference, 200 SLOC in Python and 400 SLOC in CUDA C++
We use NCCL \texttt{send} and \texttt{recv} pairs for data transmission between GPUs and socket if any side is DRAM.

We run all tests atop a single server with eight H800-80G GPUs.
We evaluate \sysname\ across four settings: (1) PD-colocated, (2) PD-colocated with caching, (3) PD-disaggregated, and (4) PD-disaggregated with caching.
The first setting runs a vanilla vLLM.
The last three settings are \sysname\ running adapted vLLM using \mempool\ APIs.
While running ShareGPT workload~\cite{sharegpt}, the PD-disaggregated with caching setting outperforms others.
Specifically, \mempool-based disaggregated inference improves JCT by up to 42\% compared to PD-colocated. Enhancing disaggregated inference with context caching can further improve JCT by 29\%!
When executing the LooGLE dataset, which features extended prompts and relatively short generation lengths, disaggregated inference boosts JCT by up to 10.8\% compared to PD-colocated setups. Additionally, context caching offers further enhancements, potentially improving JCT by 26.9\%.



In summary, we make the following contributions:
\begin{itemize}
\setlength\itemsep{0em}
\item We propose \mempool, a memory pool designed for LLM serving with a rich set of APIs.
\item We build the first disaggregated inference with context caching in \sysname\ based on \mempool\ APIs.
\item We propose a novel prompt trees-based locality-aware policy for scheduling LLM requests.

\end{itemize}

\section{Background}

\textbf{Generative LLM Inference}
LLM inference involves generating a sequence of output tokens in response to an input prompt. This process consists of two distinct phases: prefill and decode.
During the prefill phase, the model processes the prompt to generate the key-value (KV) cache. The KV cache comprises key-value pairs generated in the self-attention mechanism.
In the decode phase, the model uses the KV cache to generate tokens iteratively. The size of the KV cache grows linearly with increasing number of generated tokens.

\textbf{Inter-Request Optimization}
This type of optimization exploits dependencies among requests for better performance.
Context caching is the only known technique in this category.
To build context caching, the model stores and reuses the KV cache from the self-attention mechanism to avoid redundant computations across similar or repeated requests. This is useful in scenarios where multiple requests share common prefixes or contexts.
Two mechanisms are essential.
First, an index is required to find dependencies among requests and consequently find the preserved KV cache (see Table~\ref{tbl-mempool-index}).
Second, a modified inference engine and attention kernel to reuse the historical KV cache (see SGLang~\cite{sglang}).




\textbf{Intra-Request Optimization}
This type of optimization exploits dependencies within a request to enhance performance. Two notable examples are disaggregated inference~\cite{patel2023splitwise,zhong2024distserve,tetriserve-2024,strati2024dejavu} and sequence parallelism~\cite{infiniteLLM}.
Generally, disaggregating prefill from decode reduces interference between these two stages and allows each to scale independently with heterogeneous hardware.
However, this breaks a single request into two sub-requests and requires rigorous KV cache transmission from prefill to decode.
The same goes for sequence parallelism, in which distributed instances need to exchange the outputs of self-attention in a rigorous manner.
Overall, intra-request optimization demands efficient mechanisms for transferring the KV cache among instances.




\section{\sysname\ Overview}
{
\begin{figure*}[t]
\begin{center}
\centerline{\includegraphics[width=0.85\textwidth]{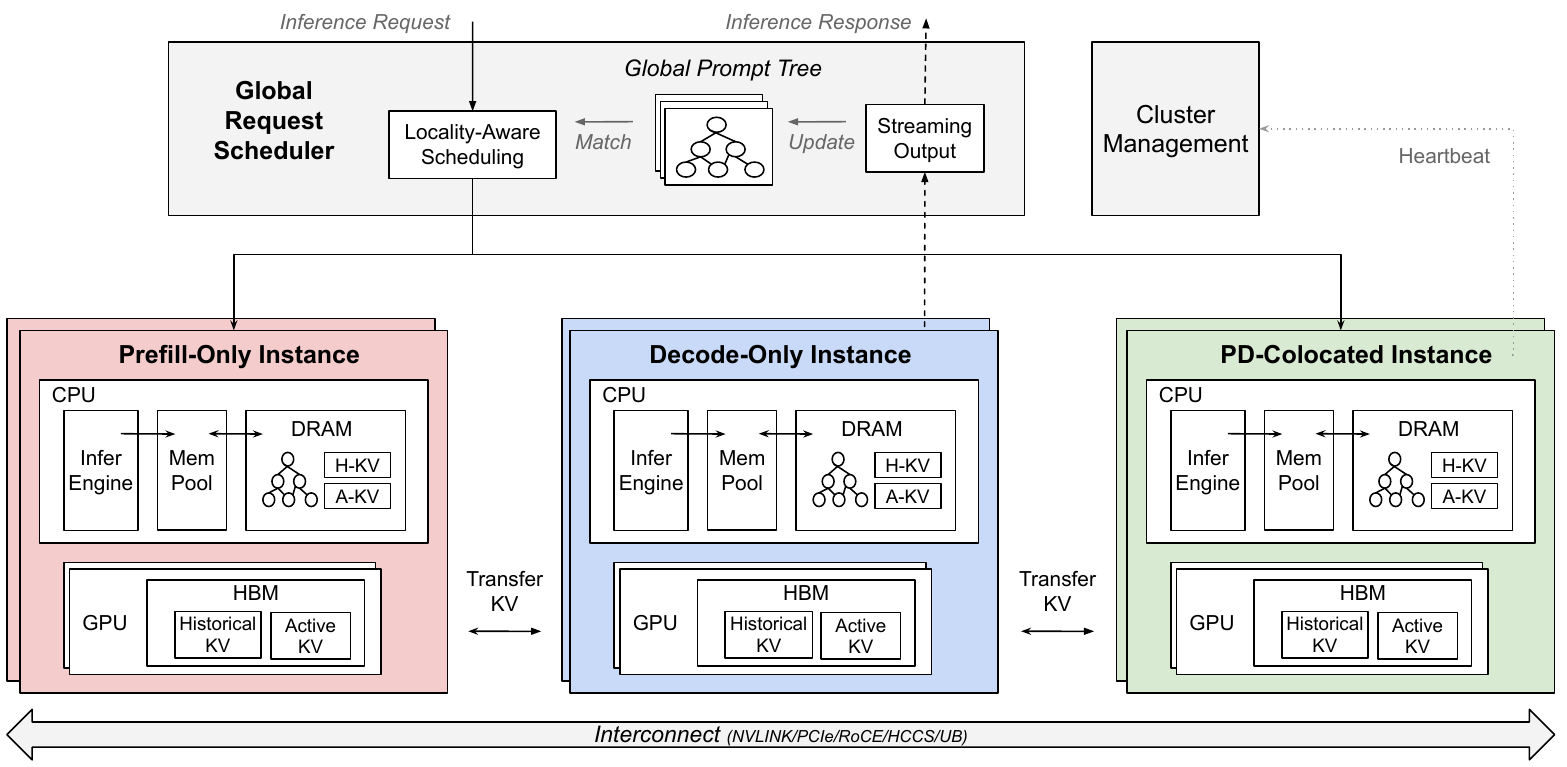}}
\caption{\sysname\ Architecture. It supports three types of inference instances: prefill-only, decode-only, and PD-colocated. Each inference engine runs over one or multiple AI servers, depending on the parallelism configuration.}
\label{fig-architecture}
\end{center}
\end{figure*}
}

\sysname\ is designed as a large-scale LLM serving system that efficiently handles inter-request and intra-request optimizations. It comprises three main components: a global scheduler, multiple types of inference instances, and an elastic memory pool (\mempool), as shown in Figure~\ref{fig-architecture}.
The \mempool\ offers a set of APIs for memory allocation, index management, and distributed transfer (\S\ref{sec-mempool}).
\sysname\ builds context caching atop both regular and disaggregated inference architectures using \mempool\ APIs (\S\ref{sec-pd-caching}).
The global scheduler forwards inference requests from users to the right inference instance. It uses locality-aware policies based on novel distributed prompt trees, maximizing KV cache reuse (\S\ref{sec-design-scheduling}).

%
%
%

%
%
%

%

\section{Elastic Memory Pool}
\label{sec-mempool}

The \mempool\ manages all memory in the inference cluster, including CPU DRAM and GPU HBM.
\mempool\ runs within each inference instance, collectively offering a set of distributed memory pool APIs (\S\ref{sec-mempool-abstraciton}).
It manages both the active KV cache used by ongoing requests and the historical KV cache retained after requests are completed. An indexing layer maps prompt tokens to the historical KV cache (\S\ref{sec-mempool-index}), ensuring efficient retrieval of cached data.
The \mempool\ has efficient mechanisms for exchanging data between instances, alleviating inference engines from dealing with heterogeneous hardware (\S\ref{sec-mempool-transfer}).
Overall, this design makes \mempool\ a versatile and generic platform capable of supporting both intra-request and inter-request optimizations within a unified system (\S\ref{sec-mempool-use-cases}).

\subsection{API}
\label{sec-mempool-abstraciton}

\begin{table}
    \caption{Elastic Memory Pool APIs. Type can be HBM-only, DRAM-only, or mixed. Each address encodes instance ID. Transfer flags can control on-demand allocation.}
    \footnotesize
    \centering
    \begin{tabular}{p{0.55in} | p{0.7in} | p{1.6in} }
    \hline
    \textbf{API} & \textbf{Parameters} & \textbf{Description} \\
    \hline
    \hline
       \texttt{alloc\_mem} &  size, type, id & alloc a certain type of memory on a given instance (@id), return addrList \\
    \hline
       \texttt{free\_mem} &  addrList & free memory \\
    \hline
       \texttt{insert} &  tokenList, addrList, flags & insert prompt token and KV cache address mapping into local index\\
    \hline
        \texttt{match} &  tokenList & find prompt's cached data if any, return addrList \\
    \hline
        \texttt{delete} &  tokenList & delete prompt's cached data if any \\
    \hline
        \texttt{swap\_out} &  num\_blocks & swap a given number of blocks from HBM to DRAM \\
    \hline
        \texttt{swap\_in} &  addrList & swap blocks with given address from DRAM to HBM \\
    \hline
        \transKV\ &  id, srcAddrList, dstAddrList, flags, private  & transfer data to the specified instance (@id), dstAddrList is optional, flags control behaviors at the destination, and private carries user data \\
    \hline
        \texttt{transfer\_ with\_insert} &  id, tokenList, srcAddrList, dstAddrList, flags, private  & transfer tokenList and its cached data to the specified instance. The receiver will call an extra insert. \\
    \hline
    \end{tabular}
    \label{tbl-mempool-api}
\end{table}

We show \mempool\ APIs in Table~\ref{tbl-mempool-api},
broadly divided into three categories: 
memory block, index, and distributed transfer.
The inference engine can use memory block APIs to allocate fixed-size memory blocks for storing KV cache or other data.
The engine can also call the index APIs for context caching solutions.
For example, once requests are finished,
the engine can call \texttt{insert} to transition the active KV cache into the historical KV cache and create a mapping from prompt tokens to the KV cache.
The engine can invoke distributed APIs, such as \transKV, to exchange the KV cache across instances when building inter-request optimizations.
Overall, the \mempool\ provides a rich framework for managing distributed memory and implementing efficient context caching and data exchange mechanisms.
%
%


\subsection{Indexing}
\label{sec-mempool-index}

\begin{table}
    \caption{Compare Indexing Methods. \sysname's \mempool\ uses prompt tokens for its generality.} 
    \footnotesize
    \centering
    \begin{tabular}{p{0.6in} | p{2in} }
    \hline
    \textbf{Indexing} & \textbf{Description} \\
    \hline
    \hline
        Token ID &  Use prompt tokens. Universally applicable.\\
    \hline
        Session ID & Use client-server session ID. Limited scope. \\
    \hline
        Document ID & Use document file ID. Limited scope.\\
    \hline
    \end{tabular}
    \label{tbl-mempool-index}
\end{table}

The \mempool\ has an index layer to map prompt tokens to the historical KV cache.
\mempool\ traverses the index whenever engines call \texttt{insert}, \texttt{match}, \texttt{delete}, etc.
The LLM serving world has three indexing methods: token, session, and document IDs (see Table~\ref{tbl-mempool-index}).
Token-based indexing is known for its generality, as it works for any shared prompt-prefix cases~\cite{sglang}.
The session and document ID indexing are simpler but can only reuse shared prompts within a chat session or across sessions using the same document~\cite{wu2024loongserve,gemini-context-caching}.
We adopt the token-based indexing method for broad applicability.
To implement this index, \mempool\ utilizes the radix tree proposed by SGLang~\cite{sglang}, with two key extensions.
First, because \mempool\ manages both GPU HBM and CPU DRAM, we enable the radix tree to reference data located anywhere in the system.
Second, since we also use the radix tree to build the global prompt tree in the global scheduler (\S\ref{sec-design-scheduling}), we add a field to indicate which inference instance holds the data.
Note that while mixed indexing methods are possible, we will explore this in future work.

To minimize data reshaping overhead, we maintain the original memory layout when transitioning the active KV cache to the historical KV cache before inserting it into \mempool. Consequently, \mempool's indexing granularity aligns with the inference engines' configuration. For example, in our tests with vLLM, which uses a block size of 16 tokens, our radix tree nodes point to KV cache blocks of 16 tokens. 



\subsection{Distributed Transfer}
\label{sec-mempool-transfer}

{
\begin{figure}[t]
\begin{center}
\centerline{\includegraphics[width=0.45\textwidth]{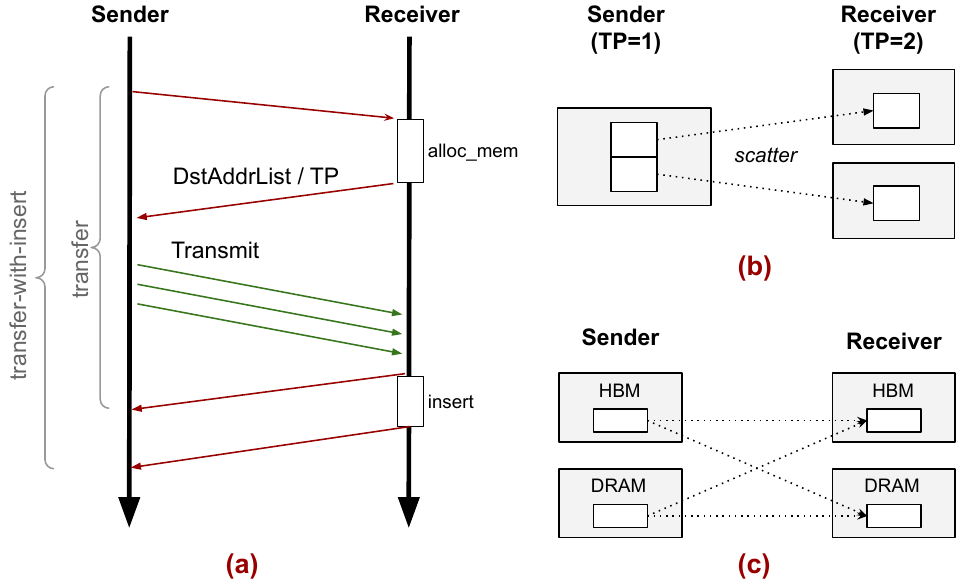}}
\caption{\mempool\ Transfer API. The left shows the workflow of \transKV\ and \transInsert. The right shows asymmetric parallelism and memory medium.}
\label{fig-mempool-transfer}
\end{center}
\end{figure}
}

The \mempool\ exposes distributed APIs for exchanging data among inference instances.
They serve as the building blocks for intra-request or inter-request dependency-exploiting techniques.
Our design rationale is to expose simple APIs that mask the underlying heterogeneity from inference engines.

Figure~\ref{fig-mempool-transfer} shows the transfer workflow and our approach to handling heterogeneity.
We break down the workflow into three steps: allocation, transmission, and insertion.
When the sender inference instance initiates a transfer, it sends a request to the receiver inference instance.
Upon receiving this request, the receiver invokes \texttt{alloc\_mem} locally to allocate HBM or DRAM based on the type specified by the sender. The receiver then returns the allocated address list and its parallelism configuration to the sender, completing the allocation step.
Then, the sender transmits the KV cache to the receiver using the fastest available path. Once all data is received, the receiver notifies the sender, completing the transmission step.
Next, The receiver checks whether this is a \transInsert\ call. If so, it invokes the \texttt{insert} function locally to insert the newly transmitted prompt tokens and historical KV cache into its local index, completing the insertion step.
Finally, the sender completes the transfer API call once the receiver returns ok.

We propose the \transInsert\ as it can avoid an extra network round-trip for establishing mapping, which is particularly useful for transferring historical KV cache from a decode-only instance to a prefill-only instance.

Additionally, users can call the \transKV\ API with a specific destination address list, allowing them to skip the initial allocation step. This feature is particularly useful for constructing layer-by-layer transmissions in disaggregated inference (see Figure~\ref{fig-mempool-pd-caching-opt}).

The transmit step is the most challenging as it must deal with three types of heterogeneities between the sender and the receiver: parallelism, memory, and network.
To manage asymmetric parallelism, the sender first checks how the KV cache is partitioned along tensor-parallel or pipeline-parallel dimensions. Once determined, the sender partitions its local cache and invokes the appropriate network primitives (top-right in Figure~\ref{fig-mempool-transfer}).
Memory asymmetry can occur if the historical KV cache has been swapped out to DRAM (bottom-right in Figure~\ref{fig-mempool-transfer}).
\mempool\ always tries to transmit data using the fastest link with the least data copies. But this is highly hardware-dependent.
If \mempool\ uses the latest hardware, such as NVIDIA SuperPOD, where all HBM and DRAM are connected by high-speed NVLINK, handling memory asymmetry is as simple as performing a memory copy.
However, on regular GPU servers, additional memory copies in the data path are inevitable.
While implementing \mempool\ distributed APIs, we find existing network primitives ill-fit for handling emerging LLM inference workloads. We will discuss this in \S\ref{sec-impl}.

\subsection{Failure Handling and Scaling}

We discuss how \mempool\ handles failures and dynamic scaling during runtime.
As Figure~\ref{fig-architecture} shows, \mempool\ is deployed as part of an inference instance. Hence, the failure and scaling granularity is a single instance, which can be one or multiple AI servers, depending on the parallelism configuration.

\sysname\ has a cluster management (CM) module as shown in Figure~\ref{fig-architecture}, which is a centralized service for maintaining cluster configuration.
The CM is responsible for adding or removing instances and monitoring cluster health.
In our current design, memory block and distributed transfer APIs can modify the states of remote instances. When an instance fails, any in-flight requests from other instances will time out. The CM detects such failures through regular heartbeats and broadcasts updated cluster information to all running instances. Upon receiving this notification, each instance releases any memory blocks allocated by the failed instance to prevent memory leaks.


\subsection{Use Cases}
\label{sec-mempool-use-cases}

{
\begin{figure}[t]
\begin{center}
\centerline{\includegraphics[width=0.45\textwidth]{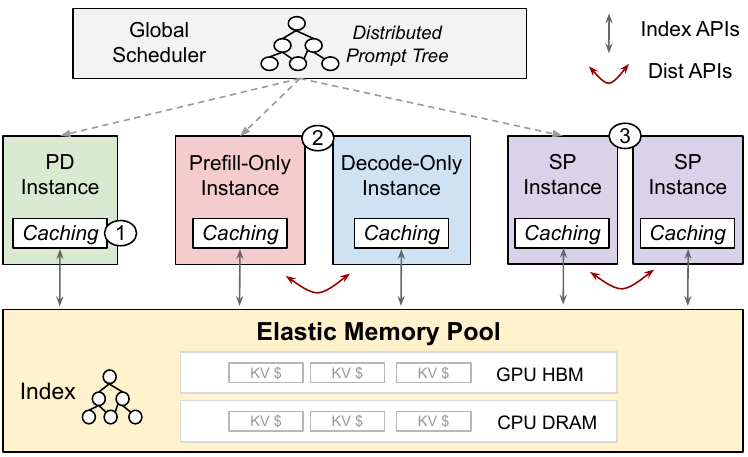}}
\caption{Use Cases Enabled By \mempool. Circle 1 is context caching. Circle 2 is disaggregated inference. Circle 3 is sequence parallelism. Solid gray lines mean \mempool\ index API calls. Solid red lines mean \mempool\ distributed APIs. \mempool\ enables all use cases in one platform.}
\label{fig-mempool-use-cases}
\end{center}
\end{figure}
}
\begin{table}
    \caption{Atomic Serving Scenarios Supported by \mempool. As a unified platform, \mempool\ supports any combo of inter-request or intra-request optimizations in one system.}
    \footnotesize
    \centering
    \begin{tabular}{c | c | l }
    \hline
    \textbf{Scenario} & \textbf{Type} & \makecell{\textbf{APIs Used}} \\
    \hline
    \hline
        \makecell{Context Caching}  & inter & index (\apiinsert,\apimatch,\apidelete,\apievict,etc) \\
    \hline
        \makecell{Disagg. Inference} & intra &\makecell{dist (\transKV,\transInsert)} \\
    \hline
        \makecell{Sequence Parallel} & intra & dist (\transKV) \\
    \hline
        \makecell{Request Migration}  & N/A & dist (\transKV) \\
    \hline
    \end{tabular}
    \label{tbl-use-case}
\end{table}

\mempool\ is a versatile and generic platform designed to support both inter-request and intra-request dependency-exploiting techniques.
Figure~\ref{fig-mempool-use-cases} illustrates how various existing inter-request and intra-request optimizations can be integrated into a unified system using \mempool\ APIs. 
Table~\ref{tbl-use-case} lists the advanced APIs employed in these optimizations.
To build context caching atop regular PD-colocated inference instances, one can use index APIs such as \apiinsert\ and \apimatch.
To build disaggregated inference, one can call \transKV\ API to send active KV cache from a prefill-only to a decode-only instance.
To build sequence-parallel (SP) inference, one can use the \transKV\ API to exchange attention outputs among instances, akin to InfiniteLLM~\cite{infiniteLLM}.

What sets \mempool\ apart is its ability to seamlessly enable these optimizations within a single system using a common set of APIs.
Next, we will showcase how to enhance disaggregated inference with context caching. We leave other combinations for future work.


\section{Caching for Disaggregated Inference}
\label{sec-pd-caching}

{
\begin{figure}[t]
\begin{center}
\centerline{\includegraphics[width=0.45\textwidth]{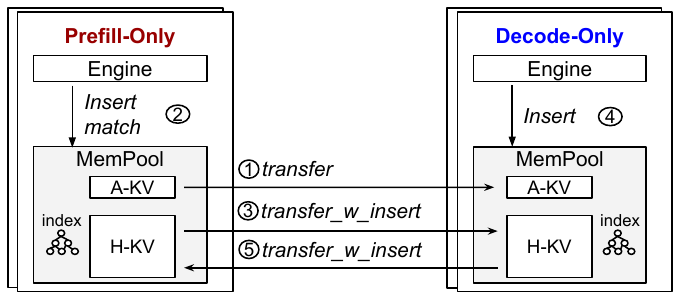}}
\caption{Enhancing Disaggregated Inference with Context Caching using \mempool\ APIs. The engine box means an adapted inference engine such as vLLM. Circled numbers mean steps taken to build the solution. A-KV is active KV cache. H-KV is historical KV cache.}
\label{fig-use-cases}
\end{center}
\end{figure}
}
\begin{table}[t]
    \caption{Towards Full-Fledged Context Caching in Disaggregated Inference. Refer to Figure~\ref{fig-use-cases} for step numbers.}
    \centering
    \footnotesize
    \begin{tabular}{l| l | l }
    \hline
    \textbf{Design} & \textbf{Steps} & \textbf{Description} \\
    \hline
    \hline
    PD-Basic & 1 & Basic PD, no caching  \\
    PD-Caching-1 & 1+2 & Caching at P \\
    PD-Caching-2 & 1+2+3+4 & Caching at D \\
    PD-Caching-3 & 1+2+3+4+5 & Full-fledged caching\\
    \hline
    \end{tabular}
    \label{tbl-mempool-pd-caching-design}
\end{table}

Context caching exploits dependency across requests,
while disaggregated inference exploits dependency within a request.
However, they fail to coexist due to missing mechanisms around KV cache management.
We enhance disaggregated inference with context caching using \mempool\ APIs. To the best of our knowledge, this is the work introducing caching to disaggregated inference.

\subsection{Design}

We show how to gradually build towards a full-fledged design in four design milestones in Table~\ref{tbl-mempool-pd-caching-design},
utilizing five key \mempool\ APIs as highlighted in Figure~\ref{fig-use-cases}.

\textbf{(a) PD-Basic.}
This is the basic disaggregated inference architecture proposed by DistServe~\cite{zhong2024distserve} and Spliwise~\cite{patel2023splitwise}.
To realize this design, 
we make minor changes to an existing inference engine (e.g., vLLM~\cite{vllm-sosp23}).
As a result, the prefill instance will call \mempool's \transKV\ API to transfer the active KV cache produced after the prefill phase to the decode instance. We carry essential metadata the decode instance requires in \transKV's private field, such as request ID, sampling parameters, prompt tokens, etc.

\textbf{(b) PD-Caching-1.}
This is the first caching-enhanced disaggregated inference design. We enable caching at the prefill-only instance by calling \apiinsert\ to retire the active KV cache as the historical KV cache such that future inferences can utilize the saved data to reduce recomputation (step 2 in Figure~\ref{fig-use-cases}).
This caching design only preserves historical KV cache produced by the prefill phase but none from the decode phase, so it works well for workloads that share long common prefix prompts, e.g., system prefix~\cite{sglang}.
The major drawback of this design is that in a multi-turn chat scenario (e.g., document QA~\cite{li2023loogle}), the prefill-only instance needs to repeatedly forward the same set of active KV cache to the decode-only instance, wasting bandwidth and affecting the time-to-second-token.
We therefore propose the next design milestone to address this issue.

\textbf{(c) PD-Caching-2.}
This design enables caching at the decode-only instance to reduce repeated data movement.
We make two key changes atop PD-Caching-1.
First, the prefill-only instance now calls \transInsert\ instead of \transKV\ such that the decode-only instance will insert the transmitted KV cache produced by the prefill phase into its local index (\S\ref{sec-mempool-transfer}).
Second, after a request finishes, the decode-only instance calls \apiinsert\ to preserve the KV cache produced by the decode phase into its local index.
With the help of locality-aware scheduling (\S\ref{sec-design-scheduling}), the prefill-only instance now only needs to transfer new KV cache data incrementally.
Though this design reduces data movement from prefill-only to decode-only instances, it does not improve context caching at the prefill-only instance since it lacks the historical KV cache from the decode phase. As a result, the benefit of context caching stays flat with increasing prompt in a multi-turn chat scenario.
We therefore propose the next design milestone to address this issue.

\textbf{(d) PD-Caching-3.}
This design enables full-fledged context caching for disaggregated inference architecture.
We make one change atop PD-Caching-2: after a request finishes, the decode-only instance calls \transInsert\ to transmit the KV cache produced by the decode phase to the prefill-only instance (step 5 in Figure~\ref{fig-use-cases}).
As a result, the prefill-only instance's preserved historical KV cache grows, and the benefit of context caching increases linearly with the number of turns.

In all, we illustrate how \mempool's simple APIs can be used to build a range of advanced solutions, from basic disaggregated inference to full-fledged context caching.
Nevertheless, \mempool\ only provides primitves for transferring data among inferences and managing local data. It is up to the users (e.g., inference engines) to decide the memory layout and the number of API calls.
Next, we will discuss design challenges around memory and network.

\subsection{Memory and Network Optimization}
{
\begin{figure}[t]
\begin{center}
\centerline{\includegraphics[width=0.4\textwidth]{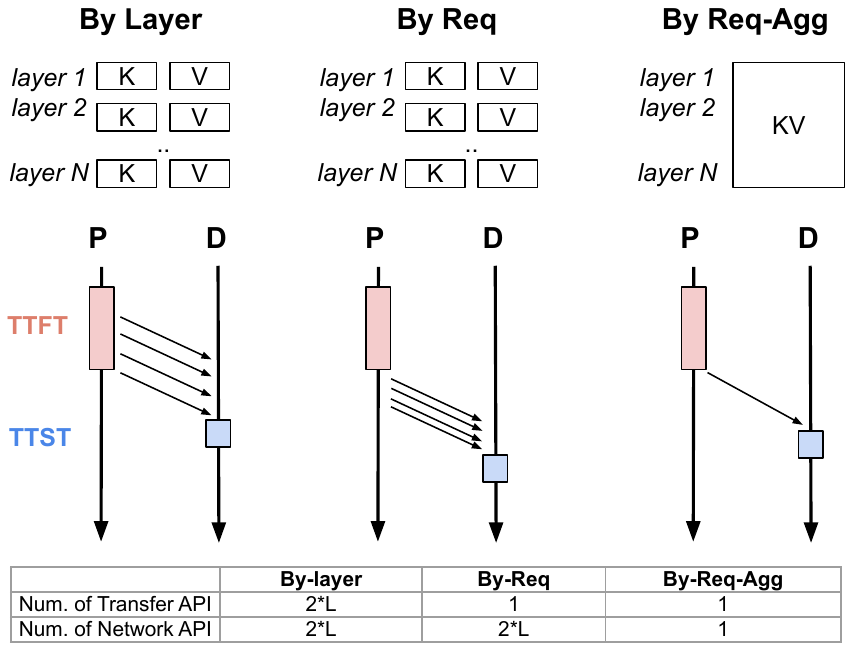}}
\caption{Optimize Network\&Memory for Disagg. Inference.}
\label{fig-mempool-pd-caching-opt}
\end{center}
\end{figure}
}

We discuss how memory and network play a key role in 
building context caching with disaggregated inference.
As Splitwise~\cite{patel2023splitwise} points out, there are two ways
to transfer active KV cache from prefill to decode instance: by-layer or by-request.
The by-layer approach transfers the KV cache once a layer has finished computation.
The by-request approach transfers the KV cache once the prefill phase is completed.
Splitwise found that by-layer outperforms by-request because it overlaps computation and communication, hence speeding up time-to-second token (or TTST).
We make the same observation when the load is low.
However, both incur non-trivial overhead with increasing load due to excessive network transfers. We find the root causes are (1) discrete memory layout and (2) inadequate network primitives.

Paging-based dynamic memory management introduced
by PagedAttention~\cite{vllm-sosp23} is now the de facto
standard in LLM serving systems,
e.g., in vLLM~\cite{vllm-sosp23}, TensorRT-LLM~\cite{tensorrt-llm}.
Regardless of where the paging mechanism is implemented (engine~\cite{vllm-sosp23} or driver~\cite{prabhu2024vattention}), the KV cache is partitioned and stored in fixed-sized memory blocks.
The block size is configurable, usually in the number of tokens, e.g., 8 tokens worth of KV cache per block.
Existing engines manage the KV cache in a fine-grained manner.
For example, vLLM allocates two blocks per LLM layer.
Given an LLM model with $L$ layers and 8 tokens per block, 
the engine needs $2*L$  blocks to store the KV cache of 8 tokens.
Although paging improves utilization~\cite{vllm-sosp23}, the discrete memory layout presents huge challenges when implementing disaggregated inference using existing AI network stacks.

The de facto network stack in AI is collective libraries such as NCCL~\cite{url-nccl}.
These libraries work best for typical AI workloads using tensor or pipeline parallelism, but they fall short in supporting LLM serving's intra-request optimizations such as disaggregated inference~\cite{tetriserve-2024} or sequence parallel~\cite{infiniteLLM}.
These new patterns require efficient point-to-point, gather, and scatter primitives between HBM or DRAM, similar to RDMA verbs~\cite{rdma-verbs}.
As discussed in \S\ref{sec-impl}, we implement \transKV\ using NCCL \texttt{send} and \texttt{recv} APIs, and each call only transmits a single block.
Since the KV cache is discrete, the number of network API calls equals the number of discrete memory blocks, regardless of whether the by-layer or by-request approach is used. This is the root cause of why both incur overhead with increasing load.

To address challenges caused by paging and poor network primitives, we propose to reduce fragmentation by aggregating smaller KV blocks into large ones, akin to using huge pages.
Specifically, instead of having two blocks per layer, we aggregate them into one block; the new block size equals $2*L$ smaller blocks. This effectively reduces the number of network API calls by $2*L$ times.
This optimization works only for the by-request approach, as the by-layer approach inevitably needs to call the network APIs at least $L$ times.
Our test shows this technique improves network performance alone by a large margin, as shown in Figure~\ref{fig-nccl-study}.

We compare by-layer, by-request, and by-request-agg (proposed optimization) in Figure~\ref{fig-mempool-pd-caching-opt} across memory layout and transmission timeline.
Our test shows that under low load, by-layer achieves the lowest JCT, but under high load, by-layer-agg outperforms by-layer thanks to reduced network calls, as shown in Figure~\ref{fig-block-aggragation}.

\subsection{Cost Model for Context Caching}

\begin{table}
    \caption{Context Caching Cost Model Factors.} 
    \centering
    \footnotesize
    \begin{tabular}{l | l }
    \hline
    \textbf{Factor} & \textbf{Description} \\
    \hline
    \hline
        Prompt-Length &  The length of the current prompt. \\
    \hline
        Cached-Ratio &  The ratio of cached tokens\\
    \hline
        Cached-Locations &  Historical KV cache locations. \\
    \hline
        Batch-Size &  The running batch size. \\
    \hline
        Block-Size &  Size of paging blocks. \\
    \hline
    \end{tabular}
    \label{tbl-cost-model}
\end{table}

In this section, we illustrate the usage and design of a cost model. 

\subsubsection{Usage of the cost model}

We propose a cost model $exec(x, y)$ that predicts the execution time for prefilling a prompt of length $x$ with a cached ratio $y$ (the percentage of the prompt being cached). This model serves two primary purposes: 1) enabling locality-aware and load-balanced global scheduling, and 2) deciding on whether to transfer KV cache or recompute them.

First, the global scheduler (GS) utilizes the cost model for locality-aware and load-balanced scheduling. Upon receiving a request, the GS matches the prompt against local prompt trees across all prefill-related instances (whether prefill-only or PD-colocated) in parallel, retrieving the corresponding cached ratios $y_p$ on each instance $p$ for this request. The cost model is then applied to predict the execution time for this request on each instance and to determine the optimal routing to instance $p$ using the following formula:
\begin{equation}
\text{argmin}_p \sum_{\text{current }x'\text{ on }p} exec(x', y'_p) + exec(x, y_p)
\end{equation}
In this equation, the first term represents the queuing delay required to process all unfinished requests on instance $p$.

Second, we use the cost model to determine whether to transfer additional cached KV cache from other instances or to proceed with computation. The GS sends the request to the chosen instance $p$ with cached ratio $y_p$. If another instance $p'$ has a larger cached ratio $y_p' \ge y_p$, we decide whether to transfer the additional KV cache (with a ratio difference $y'_p - y_p$) from $p'$ or to directly compute the prefill stage based on the following condition:
\begin{equation}
\text{transfer}(y_p, y_p') \le exec(x, y_p) - exec(x, y_p')
\end{equation}
If the condition is true, we proceed with the transfer; otherwise, we opt for recomputation. The transfer function calculates the time required by dividing the amount of transferred data by the maximum bandwidth. In practice, we apply the cost model once for a batch of requests, summing their prompt lengths and cached ratios.

\subsubsection{Design of the cost model}


The cost model $exec(x, y)$ is closely related to the transformer architecture, besides the prompt length $x$ and the cached ratio $y$. We have two approaches to estimate $exec(x, y)$. First, we collect $(x, y)$ and time-to-first-token (TTFT) pairs to develop an arch-level cost model by fitting these data pairs. Second, we collect data at the operator level, fitting functions $op(x, y)$ for various operators, with $exec(x, y)$ representing the sum of all $op(x, y)$ within the architecture.

We choose the operator-level cost model for two main reasons. First, it offers greater scalability when faced with model parallelism. Although the accuracy of predicting communication time is the same for both operator-level and architecture-level. The architecture-level cost model requires recalibration with any changes in configuration parameters, such as tensor parallelism (TP) or pipeline parallelism (PP), while the operator-level model can be readily adjusted by multiplying constants associated with PP or TP. For instance, when $TP = 2$, arch-level $exec(x, y)$ is not necessarily halved due to Amdahl's law, as the prefill stage includes both parallel and serial components. Directly halving the arch-level $exec(x, y)$, in this case, leads to a 20\% accuracy drop compared to the operator-level cost model (Figure~\ref{fig-cost-model-performance}). Second, the operator-level model is more interpretable and easier to fit. For example, operators with a complexity of $O(x^2)$ can be directly modeled using a corresponding $x^2$ polynomial function. In contrast, the arch-level cost model often requires complex machine-learning techniques for effective fitting. 

We collect profiling data and fit three types of operators: compute-bound, memory-bound, and constant operators.

\textbf{(a) Compute-bound operators}: we define $op(x,y)=(\eta-1) \cdot T_{fullwave}+T_{lastwave}$, where $\eta =\left \lceil B_{total}/SMs_{num} \right \rceil$ and $T_{fullwave}=M/N$. The $B_{total}$ represents the number of launched thread blocks and $SMs_{num}$ equals to the number of SMs in the GPU. For $T_{fullwave}$, the $M$ denotes the number of FLOP instructions in a $fullwave$ while $N$ is the maximum throughput of FLOP instruction. In practice, we can roughly define $T_{fullwave}=T_{lastwave}$ because each thread block performs the same amount of FLOP instructions and being executed in parallel. We can easily get $B_{total}$ and $M$ from the source algorithm, while only profiling one compute-bound operator to identify $N$. For instance, in the Attention-Output operator, we perform a matmul as $\left [ l,h \right ]\times  \left [ h,h \right ]$, where $l$ is the prompt length and $h$ means the hidden size of the model. If we apply tile optimization with a tile size of $t$, the number of launched thread blocks can be calculated as $b=\left \lceil l/t \right \rceil \cdot \left \lceil h/t \right \rceil$. To find the optimal $N$, we can repeatedly profile cases where $b\ge SMs_{num}$ and extend its value to other compute-bound operators.


\textbf{(b) Memory-bound operators}: Initially, we propose a formula analogous to that used for compute-bound operators, where $M$ represents memory read or write transactions and $N$ denotes the maximum throughput of read or write transactions in HBM, to calculate execution time. However, this approach is insufficient due to the multi-level cache architecture within the GPU, which complicates the accurate determination of $N$'s limits. Consequently, we instead fit the cost of memory-bound operators directly by examining the relationship between latency and the number of read and write operations. Specifically, we define $op_{attention}(x,y)$ as $ax^{2}y+bx^{2}+cx+d$ if we implement Prefix Attention using algorithm from FlashAttention-2 \cite{flashattention2-arxiv23} and obtain the values of $a, b, c, d$ through profiling.

\textbf{(c) Constant operators}: We note that these operators (normalization, activation) display a consistent execution time, which remains a fixed proportion of the total prefill time. This characteristic allows us to model their cost using a straightforward linear relationship.

\section{Locality-Aware Global Scheduling}
\label{sec-design-scheduling}

{
\begin{figure}[t]
\begin{center}
\centerline{\includegraphics[width=0.42\textwidth]{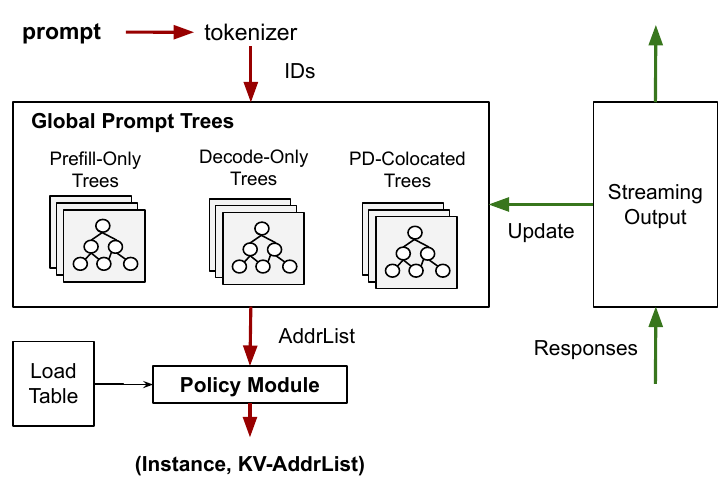}}
\caption{Global Scheduler Architecture. We highlight the global prompt trees-based locality-aware scheduling, it consists of a lookup path (left) and an update path (right).}
\label{fig-global-prompt-tree}
\end{center}
\end{figure}
}
\begin{table}
    \caption{Global Request Scheduling Policies. We compare whether they improve inter-session or intra-session context caching. A session can be an HTTP session.}
    \footnotesize
    \centering
    \begin{tabular}{c | c | c }
    \hline
    \textbf{Name} & \textbf{Intra-Session} & \textbf{Inter-Session} \\
    \hline
    \hline
        Least Load &  N & N \\
    \hline
        Session-ID-Based & Y & N \\
    \hline
        Prompt-Tree-Based &  Y & Y \\
    \hline
    \end{tabular}
    \label{tbl-gs-policies}
\end{table}

In this section, we describe \sysname's global scheduler (\GS).
The \GS\ routes requests from external services to underlying inference instances and returns generated responses in a streaming fashion.
To improve context caching at a large scale,
we propose global prompt trees and a locality-aware scheduling policy.
Figure~\ref{fig-global-prompt-tree} shows GS's architecture.

\textbf{Global Prompt Trees.}
Since \sysname\ runs three types of inferences,
the GS employs three types of prompt trees, for prefill-only, decode-only, and PD-colocated instances.
Each tree type has a set of radix trees, the same as the ones used by \mempool\ with one extra field per tree node pointing to the instance storing the KV cache.
The global prompt trees support regular \apiinsert\ and \apimatch\ APIs as listed in Table~\ref{tbl-mempool-api}.
For now, both \gs's global prompt tree and each inference's local prompt tree share the same indexing granularity.

\textbf{Scheduling.}
When a request arrives at the GS, it goes through the following steps.
First, the \GS\ runs a tokenizer to turn prompt strings into token IDs.
Second, the \GS\ queries the global prompt tree by concurrently calling \apimatch\ against all types of trees.
Third, the \GS\ sends query results along with current load info to a policy module.
The policy module chooses an instance with the longest common prefix (i.e., the largest preserved historical KV cache).
Once the instance is chosen, the \GS\ checks whether there exist instances storing extra historical KV cache that is not present in the chosen instance. If so, the policy engine also outputs a list of such instances and the corresponding token IDs.
Finally, the \GS\ sends the request and metadata to the chosen instance.
We update the global prompt trees when instances return responses back to callers.

\textbf{Discussion.}
(1) Our proposed prompt-tree-based policy is a \textit{best-effort} scheduling policy. It tries to maximize context caching reusing opportunities. Since the \GS\ only updates its prompt tree when responses pass through it, the \GS\ is unaware of local eviction events in underly instances. Therefore, the \GS's local prompt tree can be outdated. We address this issue by configuring the global prompt trees with a time-to-live (TTL), commonly in minutes.
(2) We compare three global request scheduling policies in Table~\ref{tbl-gs-policies} across two dimensions.
The least-load policy means selecting the instance with the least load, unaware of any locality.
The session-ID-based policy routes requests based on a connection ID (e.g., HTTP session). This policy enables context caching within a session.
Our prompt-tree-based policy can exploit caching opportunities across sessions, reusing most context caching.

\section{\sysname\ Implementation}
\label{sec-impl}

\sysname\ has three key parts: \mempool, context caching with disaggregated inference, and global scheduler.
We implement \mempool\ from scratch,  5K SLOC in Python and   1.6K SLOC in C++.
We modify vLLM~\cite{vllm-sosp23} to build context caching with disaggregated inference, 200 SLOC in Python and 400 SLOC in CUDA C++.
The global scheduler and the cluster management have 600 SLOC in Python.

\textbf{\mempool.}
It has two parts: a Python-based library that exposes API to the inference engine and a C++ core part that executes data transmission.
Currently, \mempool\ uses NCCL's \texttt{send} and \texttt{recv} point-to-point APIs to transmit data between HBM and uses socket API if any side contains DRAM. We have not implemented RDMA-based transmission because we only have a single AI machine. 
As we've mentioned earlier, NCCL is a collective library designed for typical tensor and pipeline parallel AI workloads but not ideal for point-to-point communication.
Specifically, its \texttt{send} and \texttt{recv} APIs only specify source addresses but no destination addresses. Hence, ensuring ordering between a sender and a receiver is challenging, especially if we aim to achieve high parallelism using multiple threads.
As a result, we end up using a single thread per NCCL communicator to ensure ordering.
Additionally, as NCCL has no gather or scatter APIs, we call the send-recv API pair multiple times to transmit data across heterogeneous parallelism instances (Figure~\ref{fig-mempool-transfer}).

\textbf{Context Caching with Disaggregated Inference.}
We adapt vLLM~\cite{vllm-sosp23} to using \mempool\ APIs.
Specifically, we replace its original cache engine and hash-based prefix caching with \mempool.
To realize block aggregation, we modify several CUDA kernels such as the paged\_attention, swap\_blocks, reshape\_and\_cache.


%

\section{Evaluation}

\subsection{Setup}

We describe the physical server, baseline systems, and LLM model used in our evaluation. 

\textbf{Server.}
We run all tests on a single NVIDIA DGX H800 server.
It has 8 H800-80GB GPUs interconnected by NVLink (400GB/s bandwidth).
It has 192-core Intel Xeon Platinum CPUs @2.4 GHz and 2 TB DRAM.
We use Ubuntu 20.04 with Linux kernel 5.16.7 and CUDA 12.2.

\textbf{Baseline.}
We use vLLM-0.4.0 as our baseline for running PD-colocated instances.
\sysname\ uses the version for building context caching enhanced disaggregated inference.

\textbf{Model.}
We use Llama2-13B with tensor parallel (TP) configured as 2 for all our tests.
We use this model size mainly because it allows us to create four inference instances within a single server; a larger model would lead to fewer instances.

\textbf{Metrics.}
For end-to-end benchmarking, we report the following metrics: time-to-first-token (TTFT), job completion time (JCT), and time-per-output-token (TPOT).

\subsection{Workloads}

\begin{table}
    \caption{Workloads Used in Our Work.}
    \footnotesize
    \centering
    \begin{tabular}{l | l | l}
    \hline
    \textbf{Type} & \textbf{Workload} & \textbf{Description} \\
    \hline
    \hline
        Chat & ShareGPT & Chat history with ChatGPT \\
        QA & LooGLE & Long document QA \\
        Agent & ReAct & Agent with acting \& reasoning \\
    \hline
    \end{tabular}
    \label{tbl-workloads}
\end{table}
{
\begin{figure}[t]
\begin{center}
\centerline{\includegraphics[width=0.45\textwidth]{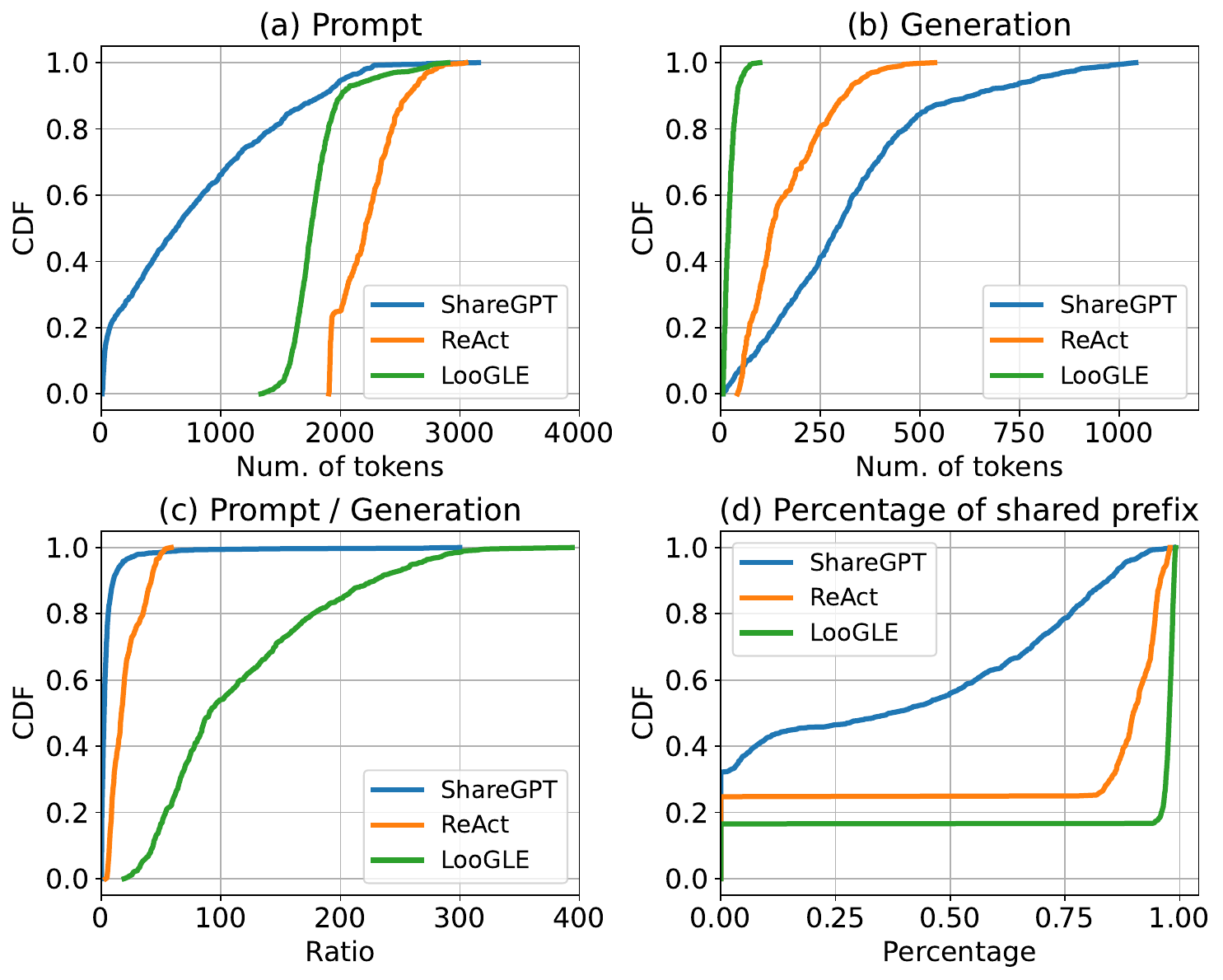}}
\caption{Workload Statistics. (a) Prompt length. (b) generation length. (c) The ratio of prompt-len over generated-len. (d) The percentage of the shared prefix in each workload.}
\label{fig-prompt-gen-tokens-len-distribution}
\end{center}
\end{figure}
}

We describe workloads used for our end-to-end tests.
We use three representative workloads as listed in Table~\ref{tbl-workloads}.
(1) ShareGPT~\cite{sharegpt}
is a real-world dataset containing user-shared ChatGPT conversations.
%
Requests from the same conversation form a session and share causal dependencies: clients send a request to the system only after they receive the response to the conversation’s previous request.
We will replace the given response with the LLM-generated content.
(2) LooGLE~\cite{li2023loogle}
is an evaluation benchmark for LLM long context understanding, containing long documents with QAs.
Similar to ShareGPT, requests constructed from the same document form a session and share causal dependencies. 
(3) ReAct~\cite{yao2022react}
is an agent acting and reasoning framework.
We use traces generated from running the ReAct agent with HotpotQA~\cite{yang2018hotpotqa} dataset.
%

\textbf{Workload Statistics}.
We study the above workloads in Figure~\ref{fig-prompt-gen-tokens-len-distribution}.
%
We show the distribution of prompt and generated token length, their ratio, and the percentage of shared prefixes representing potential context caching benefits.
Generally, ShareGPT exhibits uniform distribution across all four dimensions.
LooGLE has long prompts, short generation lengths, and a large percentage of shared prefixes because each request has a long document in its prompt.
Since the document exceeds our model's context window, we only take the first 1k tokens of the document and keep the first five associated questions.
ReAct also has long prompts and a large percentage of shared prefixes because each request has a long two-shot example in its prompt.
Unlike LooGLE, requests from ReAct have relatively long generation lengths because of the long and thorough reasoning and actions generated from LLMs. 

\textbf{Arrival Pattern.}
None of the above workloads has arrival patterns.
We simulate a request’s arrival time by sampling it from a Poisson distribution under different request rates.
We maintain the causal dependency for requests belonging to the same session: a request is only sent to the system after receiving the response to the session’s previous request. 


\subsection{End-to-End Applications}

{
\begin{figure*}[t]
\begin{center}
\centerline{\includegraphics[width=\textwidth]{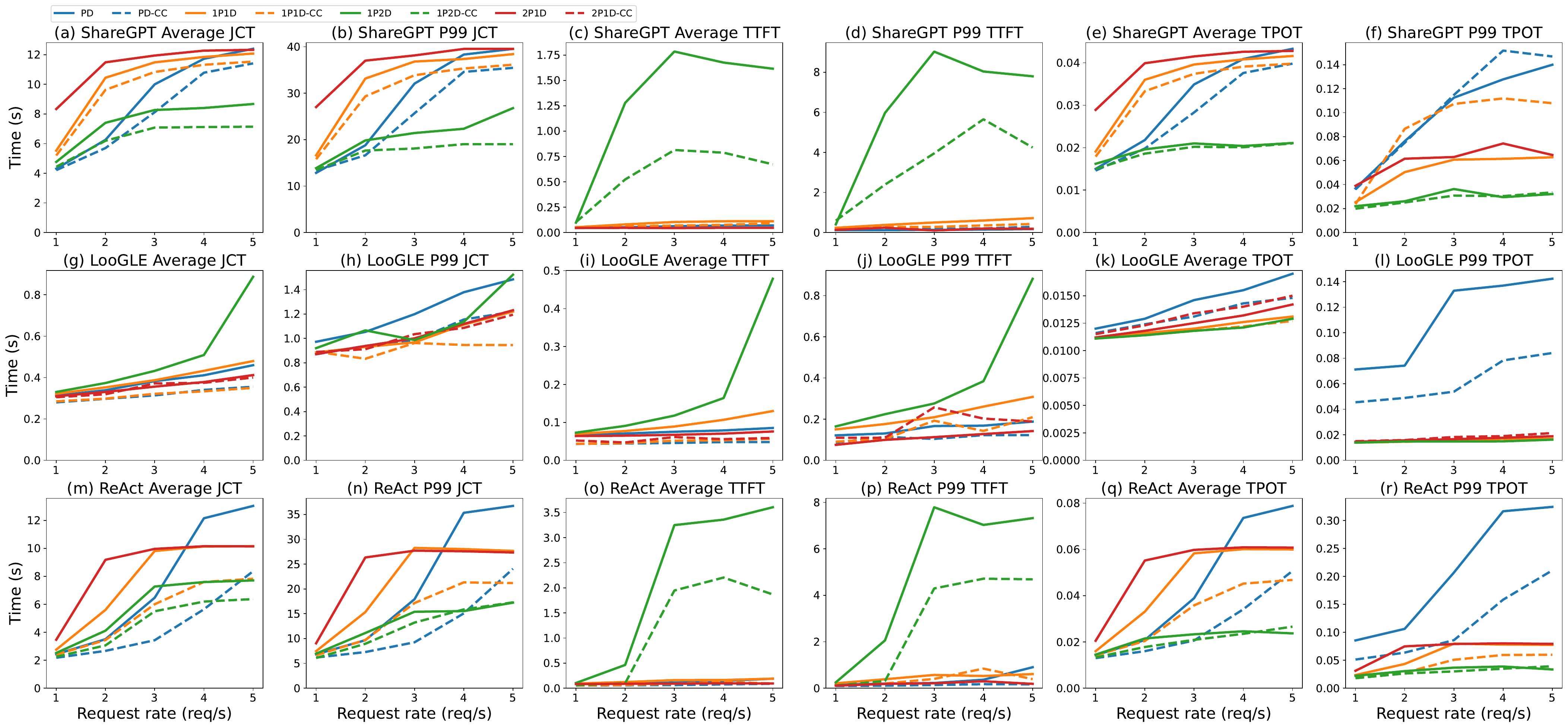}}
\caption{End-to-End Evalution. The x-axis is the request rate per inference instance, 1P1D counts as two instances.}
\label{fig-e2e-results}
\end{center}
\end{figure*}
}

We study the benefits of context caching, disaggregated inference, and when both are combined.
We create four different settings:
(1) PD denotes PD-colocated.
(2) PD-CC denotes PD-colocated with context caching.
(3) 1P1D denotes disaggregated inference with a single prefill-only and a single decode-only instance. The numbers can vary.
(4) 1P1D-CC denotes 1P1D with context caching (PC-caching-3).
Note that PD-colocated runs vanilla vLLM.
The other three settings are run with \sysname.
The request rate is calculated per instance.
Assume a 5 req/s rate, then a 1P1D setup will take 10 req/s.
We ensure an equal number of instances in all tests.
Also, we use prompt-tree-based scheduling and memory aggregation (by-req-agg). All results are in Figure~\ref{fig-e2e-results}.

\textbf{ShareGPT}.
Compared to PD-colocated,
disaggregated inference (1P2D over PD) improves average and P99 JCT by 30\% and 42\%, respectively.
Enhancing disaggregated inference (e.g., 1P2D) with context caching (e.g., 1P2D-CC) further improves average and P99 JCT by 17\% and 29\%, respectively.
It also improves average and P99 TTFT by 58\% and 45\%, respectively.
Since ShareGPT has the longest generation length of all three workloads, compared to 2P1D, 1P2D improves JCT because it improves TPOT but at the cost of heavily loaded prefill instances, hurting TTFT.

\textbf{LooGLE and ReAct.}
Both have long prompts and relatively short generation lengths. 
%
%
For LooGLE, disaggregated inference improves average and P99 JCT by 10.3\% and 10.8\%, respectively.
Context caching further improves average and P99 JCT by 26.9\% and 22.5\%, average and P99 TTFT by 56.2\% and 45.2\%.
For ReAct, disaggregation increases average and P99 JCT by 40.8\% and 53.1\%. Caching further enhances these metrics by 26.7\% and 21.4\%, and average and P99 TTFT by 78.5\% and 84.9\%.

\subsection{Microbenchmarks}

{
\begin{figure}[t]
\begin{center}
\centerline{\includegraphics[width=0.45\textwidth]{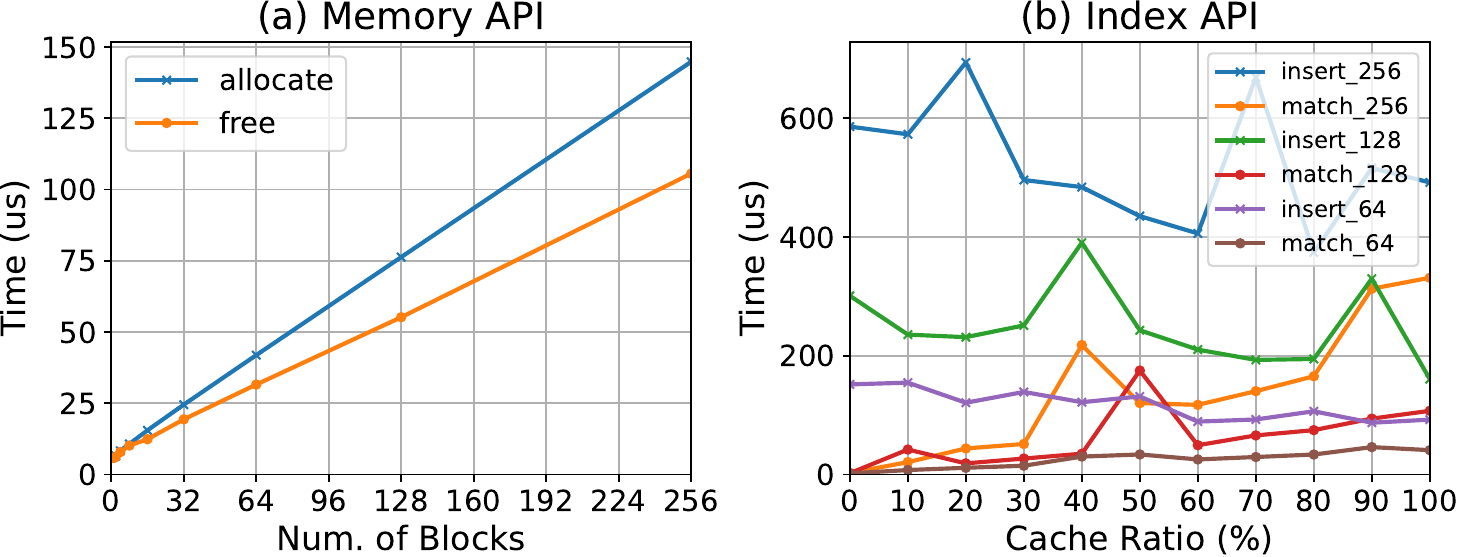}}
\caption{\mempool\ API Study. (a) The latency of Memory APIs with varied numbers of blocks. (a) The latency of key Index APIs with varied cache ratio and number of blocks. }
\label{fig-api-study}
\end{center}
\end{figure}
}
{
\begin{figure}[t]
\begin{center}
\centerline{\includegraphics[width=0.45\textwidth]{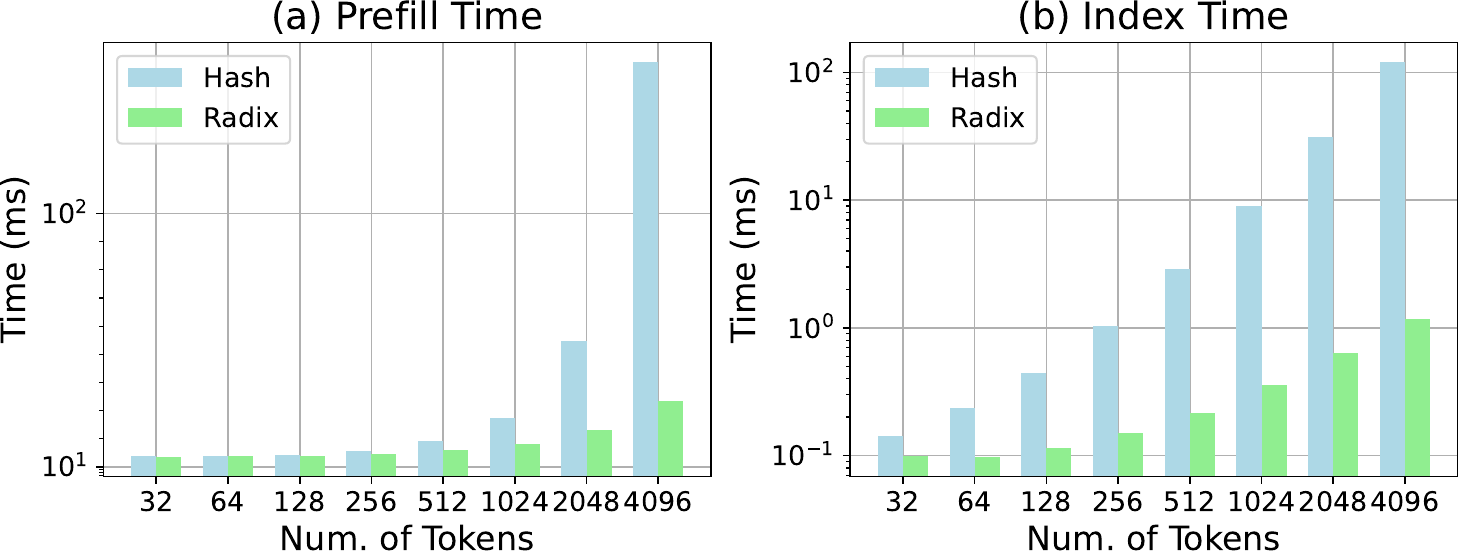}}
\caption{Caching Study. PD-Colocated. Hash is vanilla vLLM. Radix is an adapted vLLM with \mempool.}
\label{fig-comp-hash-radix}
\end{center}
\end{figure}
}
{
\begin{figure}[t]
\begin{center}
\centerline{\includegraphics[width=0.45\textwidth]{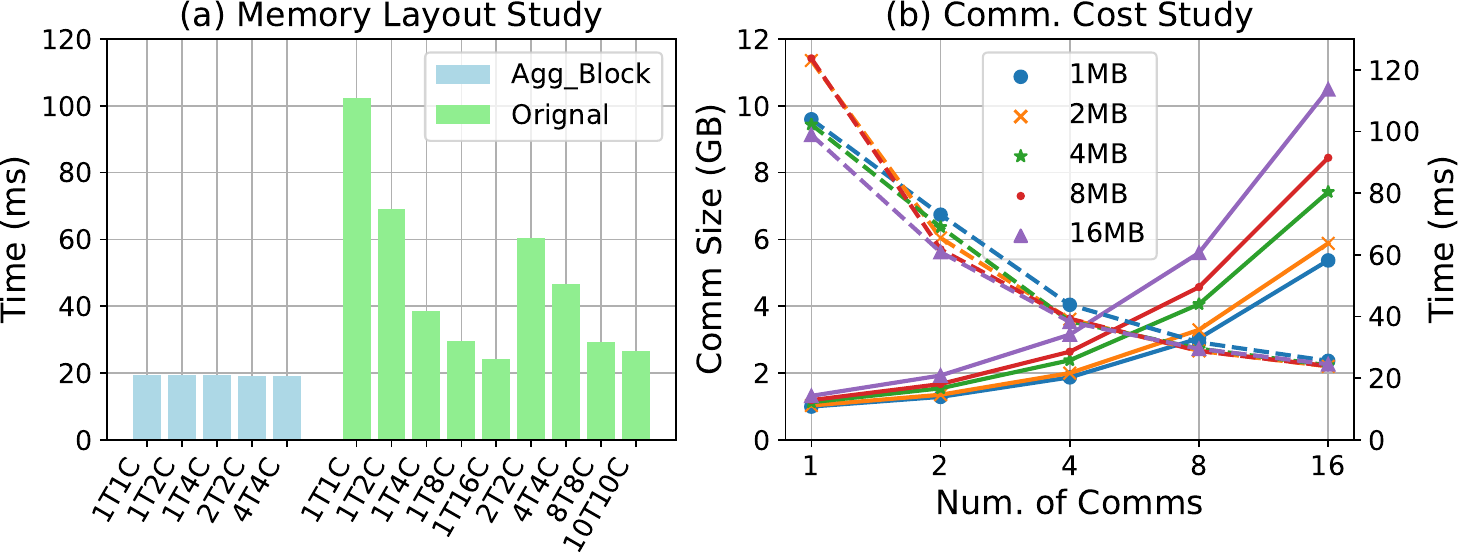}}
\caption{Network and Memory Layout Optimization Study. T is short for threads. C is short for NCCL communicators. The right figure compares the performance and HBM usage with varied NCCL buffer sizes. The default is 4 MB.}
\label{fig-nccl-study}
\end{center}
\end{figure}
}

\textbf{\mempool\ API Study.}
We first study the main \mempool\ APIs in Figure~\ref{fig-api-study}.
Without loss of generality, we show a few key APIs.
Memory APIs' latency increases linearly with the number of blocks, taking roughly 800 ns per block.
For index APIs, we mainly run insert and match. We vary the number of blocks. A 256 block equals 4K tokens. The latency mostly stays flat with an increasing cached ratio. It takes at most 0.7 ms to insert a 4K prompt. 
In all, \mempool\ APIs are lightweight and fast.

\textbf{\mempool\ Caching Study.}
We compare vanilla vLLM's hash-based index with \mempool's radix-based index.
We run both on a PD-colocated instance with no cached data.
We record the prefill time, which consists of two parts: check index and model forward. Figure~\ref{fig-comp-hash-radix} shows that vanilla vLLM's hash-based prefix mechanism incurs a huge overhead as the prompt length increases. 
In all, using \mempool\ for basic context caching incurs minimal overhead.

\textbf{Block Aggregation Study.}
We study how the proposed memory aggregation helps.
We compare two settings: (1) original discrete memory layout (Original) and (2) proposed aggregated memory layout (Agg\_Block).
The test transmits the KV cache generated from a 2048-token prompt.
We tune several key NCCL parameters: communicator, stream, buffer size, and threads.
Figure~\ref{fig-nccl-study} presents the results.
First, the aggregation method outperforms the vanilla memory layer by a large margin.
Second, a single communicator is enough when the memory block is large.
When the memory block is smaller, multiple communicators are required for better performance, but as the right figure shows, increasing the number of communicators consumes extra HBM.

\textbf{By-Req-Agg Study.}
{
\begin{figure}[t]
\begin{center}
\centerline{\includegraphics[width=0.4\textwidth]{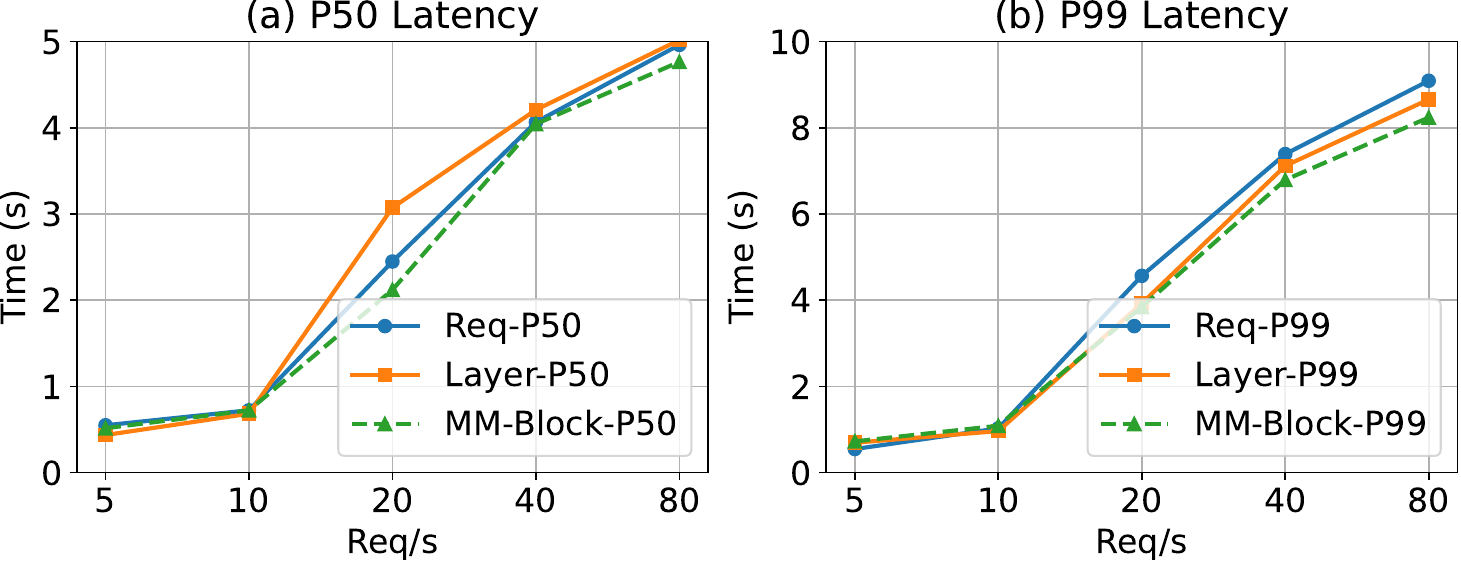}}
\caption{Compare By-Layer, By-Req, and By-Req-Agg.}
\label{fig-block-aggragation}
\end{center}
\end{figure}
}
We run a 1024-prompt-32-decode workload to understand these mechanisms.
We vary the request rate and show results in Figure~\ref{fig-block-aggragation}. The proposed by-req-agg outperforms both by-layer and by-req.

\textbf{Context Caching Study.}
{
\begin{figure}[t]
\begin{center}
\centerline{\includegraphics[width=0.45\textwidth]{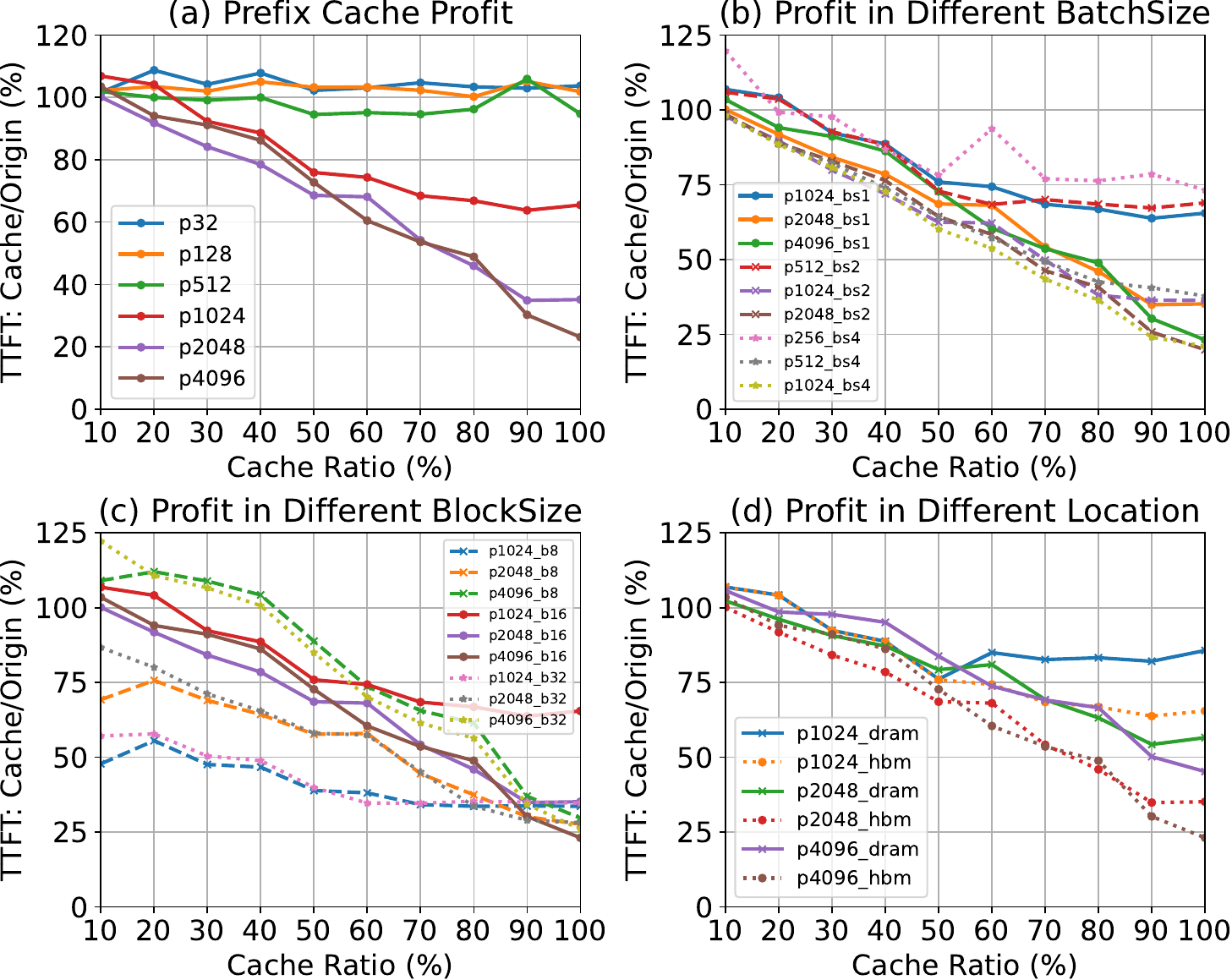}}
\caption{Context Caching Cost Model. All figures have cached-ratio has the x-axis. Each line represents a different prompt length. All y-axis represent the TTFT improvement over the no-caching case.
(a) studies the prompt-len factor.
(b) studies the batch-size factor.
(c) studies the block-size factor.
(d) studies the cached-location factor.
}
\label{fig-cost-model}
\end{center}
\end{figure}
}
Figure~\ref{fig-cost-model} presents the result with several key takeaways.
(1) The benefit of caching improves with a larger cached-ratio.
(2) For the same cached-ratio, longer prompts have higher improvement.
(3) Batch size effectively translates to prompt length. Hence, we need to consider batch size along with cached-ratio.
(4) When the historical KV cache data is located in DRAM, we must swap it into HBM before using it during prefill. Yet, the benefit of reducing computation largely offsets the cost of data movement. Regardless of where data is located, TTFT improves once the cached-ratio exceeds a certain threshold.

\textbf{Cost Model Study.}
{
\begin{figure}[t]
\begin{center}
\centerline{\includegraphics[width=0.45\textwidth]{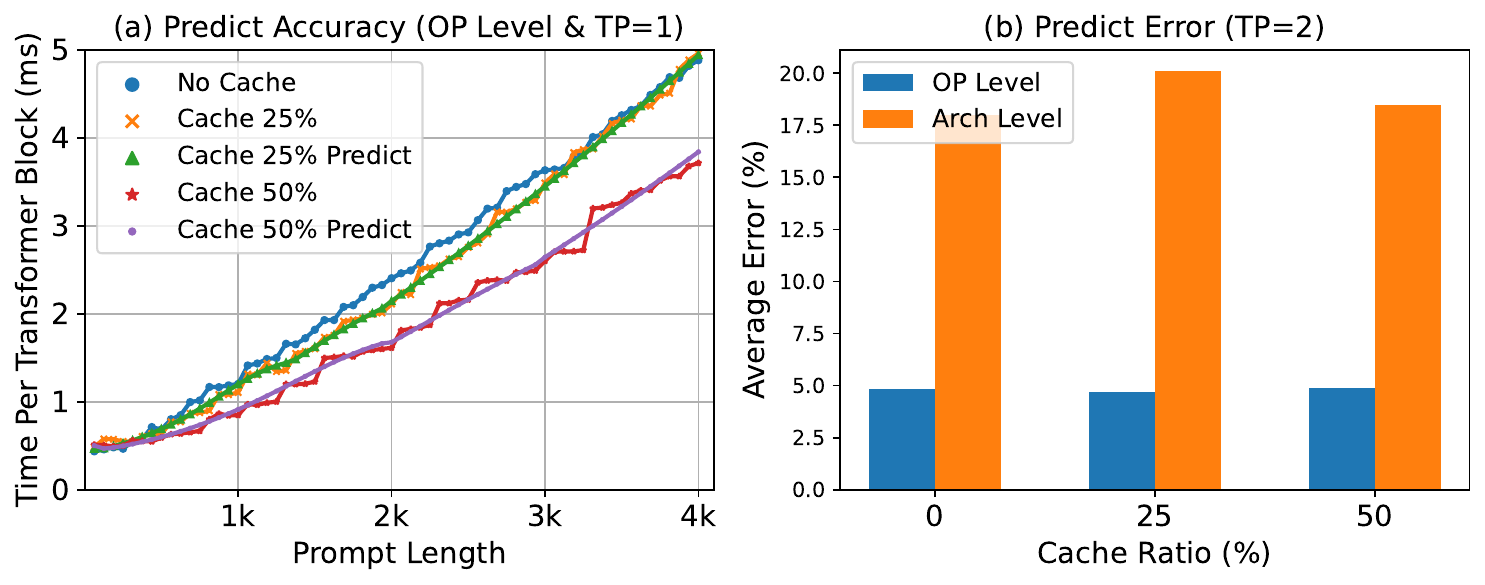}}
\caption{Cost Model Accuracy.
(a) Operator-Level Model Accuracy.
(b) Compare Operator-Level and Arch-Level.
}
\label{fig-cost-model-performance}
\end{center}
\end{figure}
}
We compare the operator-level and arch-level cost models from two aspects: precision and scalability. We model in both cached and uncached cases, recording the prefill times in different Tensor Parallism. Figure~\ref{fig-cost-model-performance} shows that compared with the arch-level cost model, the operator-level not only has a better prediction error but also shows better performance in scalability.

\textbf{Global Scheduler Study.}
{
\begin{figure}[t]
\begin{center}
\centerline{\includegraphics[width=0.45\textwidth]{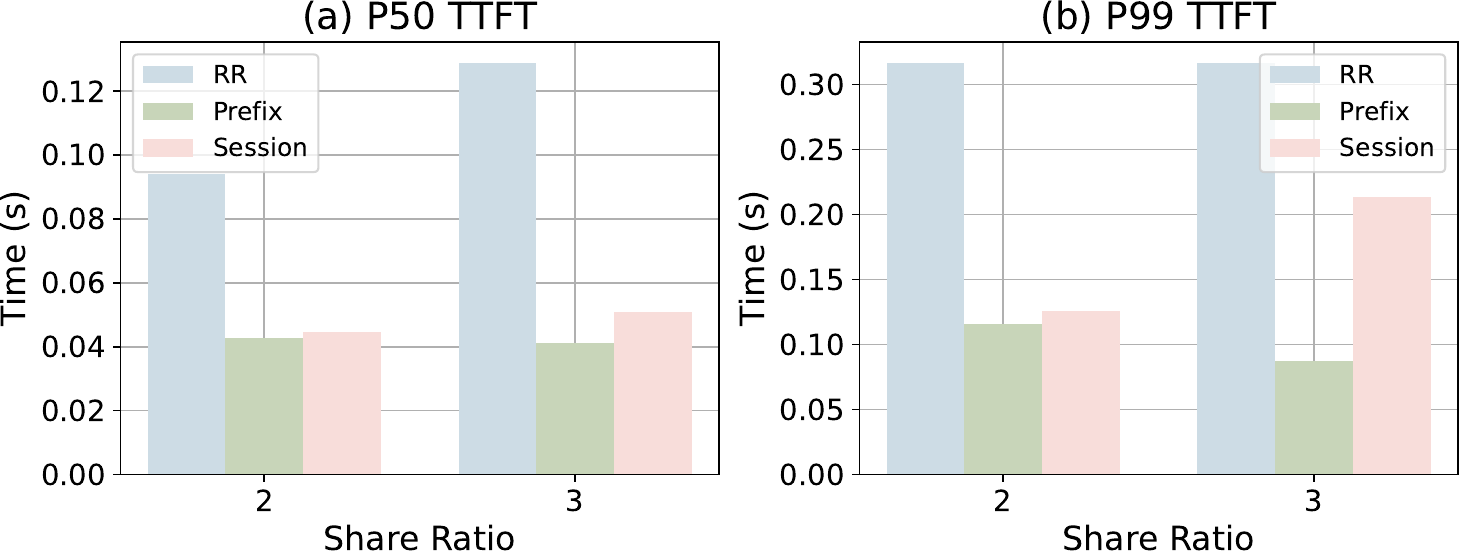}}
\caption{Global Scheduler Policy. Share Ratio represents the ratio of the number of identical requests.
}
\label{fig-gs-policy}
\end{center}
\end{figure}
}
We compare policies as listed in Table~\ref{tbl-gs-policies}. We selected 80 sessions from LooGLE, roughly 250 requests. We propose a share ratio. A share ratio of 2 means duplicating this set of sessions. While running a 3P1D setup, Figure~\ref{fig-gs-policy} shows that compared to intra-session scheduling, prompt-tree-based scheduling improves P99 TTFT by 59\% since it maximizes KV cache reuse.
\section{Related Work}

Our work is unique in proposing a standalone \mempool\ module and developing a holistic serving system \sysname\ using \mempool\ APIs.

\textbf{Disaggregated Inference.}
Four papers propose the disaggregated inference idea concurrently within a short 3-month span: Splitwise~\cite{patel2023splitwise}, TetriServe~\cite{tetriserve-2024}, DistServe~\cite{zhong2024distserve}, and Dejavu~\cite{strati2024dejavu}.
Generally, disaggregating prefill from decode reduces interference between these two stages and allows each to scale independently with heterogeneous hardware.
More recently, LoongServe~\cite{wu2024loongserve} takes a step further by enabling dynamic scaling.
All prior work builds disaggregated inference by modifying the inference engine in an ad-hoc manner.
Our work takes a different approach by first abstracting out the \mempool\ component and then building disaggregated inference as a use case of \mempool.

\textbf{Context Caching.}
Caching reduces recomputation, hence reducing TTFT and improving throughput.
The benefits of context caching are well-studied in Pensieve~\cite{pensieve}, Cache Gen~\cite{liu2023cachegen}, SGLang~\cite{sglang}, and Prompt Cache~\cite{gim2024promptcache}.
More recently, Google started a commercial offering of context caching for their Gemini models~\cite{gemini-context-caching}.
All prior work builds context caching in a PD-colocated setup. Using \mempool\ APIs, we take a step-by-step approach to building the first-ever context caching solution atop disaggregated inference.

\textbf{Scheduling.}
Scheduling plays a key role in improving serving efficiency.
%
At the local layer:
Orca~\cite{yu2022orca} proposes iterative-level scheduling to reduce bubbles.
Sarathi~\cite{sarathi-arvix23, sarathi2} proposes chunked-prefill to overcome suboptimal prefill processing.
FastServe\cite{fastserve-arxiv23} utilizes a multi-level priority feedback queue to minimize JCT.
At the global layer:
MuxServe~\cite{duan2024muxserve} formulates a multiplexing problem and proposes a novel placement algorithm and adaptive batch scheduling strategy to identify optimal colocations in LLM serving.


\textbf{Generic Memory Optimization}.
Many works try to optimize memory usage. For example, using quantization\cite{rethinking-icml20,smoothquant-icml23,Llmint8-arvix23, flexgen-pmlr23, PTQ-arixv23, gptzip-icml23, spqr-arvix22, optq-eiclr22} to compress the model weights into lower precision, using paging to reduce fragmentation~\cite{vllm-sosp23}, and low-level algorithm and kernel optimizations~\cite{zeroquant-nips22, rammer-osdi20, lightseq-arvix20, flashattention-nips22, flashattention2-arxiv23, flashdecoding++-2023, bytetransformer-ipdps23}. We refer readers to ~\cite{zeng2024cap-llm,zhou2024survey-llm} for more details.

\section{Conclusion}

In this paper, we presented \sysname, a novel system designed to enhance the efficiency of LLM serving by unifying inter-request and intra-request optimizations.
The core of \sysname\ is a distributed \mempool\ that manages KV caches across distributed instances.
\sysname\ builds context caching, disaggregated inference, and their combo using \mempool\ APIs.
End-to-end results show \sysname\ can improve JCT, TTFT, TPOT by a large margin.



\bibliographystyle{plain}
\bibliography{paper}

\end{document}